# A Platform for Spreadsheet Composition

Pierpaolo Baglietto [(1)], Martino Fornasa [(1)], Simone Mangiante [(1)],
Massimo Maresca [(1)], Andrea Parodi [(2)], Michele Stecca [(1)]

[(1)] Computer Platform Research Center (CIPI) - University of Padova, University of
Genova, Italy
[(2)] M3S S.r.l. – Via Al Molo Cagni, Genova, Italy

p.baglietto@cipi.unige.it, m.fornasa@cipi.unige.it, simone.mangiante@cipi.unige.it,
m.maresca@cipi.unige.it, a.parodi@m3s.it, m.stecca@cipi.unige.it

## ABSTRACT

*A huge amount of data is everyday managed in large organizations in many critical business
sectors with the support of spreadsheet applications. The process of elaborating spreadsheet data
is often performed in a distributed, collaborative way, where many actors enter data belonging to
their local business domain to contribute to a global business view. The manual fusion of such
data may lead to errors in copy-paste operations, loss of alignment and coherency due to multiple
spreadsheet copies in circulation, as well as loss of data due to broken cross-spreadsheet links. In
this paper we describe a methodology, based on a Spreadsheet Composition Platform, which
greatly reduces these risks. The proposed platform seamlessly integrates the distributed
spreadsheet elaboration, supports the commonly known spreadsheet tools for data processing and
helps organizations to adopt a more controlled and secure environment for data fusion.*

## 1 RISKS IN DISTRIBUTED SPREADSHEET COLLABORATIVE DATA PROCESSING

Companies and organizations in many critical business areas base many of their most
critical data processing tasks on spreadsheet applications [1][2]. The reasons why this
happens have been investigated and researched by the European Spreadsheet Risks
Interest Group, as well as many of the potential risks associated with the use of
spreadsheets in these processes [3][4].

Some of these risks are due to bad practices or lack of control in the development phase
of the spreadsheets by the End Users [5], but many risks are also hidden in the run-time
processing of data contained in distributed spreadsheets [6]. In many large organizations
spreadsheet-based data processing often takes place in a distributed, collaborative way,
where many actors produce or consume spreadsheet-based data and contribute to the
setup of a real spreadsheet-based "distributed workflow".

The propagation and the update of data in this workflow is commonly performed either
using the tools provided by spreadsheet applications (like cross-spreadsheet links) or
using manual techniques, like copy and paste operations over spreadsheets. Access to
spreadsheet data is also not controlled, allowing different methods like concurrent access
by different actors to a shared location or circulation of spreadsheets by email. The
combinations of these practices may lead to errors and/or inefficiencies, the most
common of which can be summarized as follows:

- inconsistent cross-spreadsheet links due to uncontrolled delete/rename/replace
  operations of spreadsheet files potentially leads to data loss;

- proliferation of many copies of data due to spreadsheet circulation by email or other means leads to data replication, lack of control on spreadsheet versions and potential usage of not up-to-date data;
- uncontrolled concurrent access to a shared spreadsheet for manual data update leads both to high latency time for update (only one actor at time is allowed to modify data) and to uncontrolled changes in the spreadsheet data;
- manual copy/paste operations for propagating data from one spreadsheet to another leads to mistyping errors and violations of cell formats.

In order to provide a more controlled approach to distributed spreadsheet processing, we investigated the application of Data and Services Composition Platform [7] in several scenarios where spreadsheet-based data processing plays a major role. This investigation led both to the definition of an approach for applying data composition to the spreadsheet case and to the development of a prototype platform supporting this methodology. We will discuss why such approach and platform can help to reduce the risks described above and improve distributed spreadsheet processing.

The presentation is organized as follows. Section 2 introduces the proposal through an example. Section 3 describes the platform features and architecture, and provides implementation details. Section 4 discusses other approaches to the problem and presents a comparison with such approaches. Section 5 provides some concluding remarks.

## 2    INTRODUCTION THROUGH A REAL CASE EXAMPLE

### 2.1    A Real Case Scenario

A car selling enterprise is organized in a hierarchical structure: a certain number of Car Dealer entities (CDs) located in the same area are managed by an Area Sales Manager entity (ASM). The ASM needs to collect sales information from the CDs in order to evaluate them through a performance index. This performance index is the percentage of completion of a sales target assigned to a particular CD by the ASM: every CD must know only his own target whereas the ASM wants to publish a comparison of CDs based on the performance index (see Figure 1).

We suppose that:

- a Virtual Private Network exists among CDs and ASM, we call it Enterprise Domain Network;
- all the CDs keep track of their sales report on their local spreadsheets;
- the ASM calculates performance indexes using his own complex local spreadsheet;
- the ASM publishes the comparison to the entire enterprise through a new spreadsheet.

Every CD can view the performance index based comparison in any moment and it should be updated to the last car sold by any CD: when a CD insert a new sold car in its local spreadsheet, its performance index must change immediately, even though the ASM is not working on its spreadsheet.

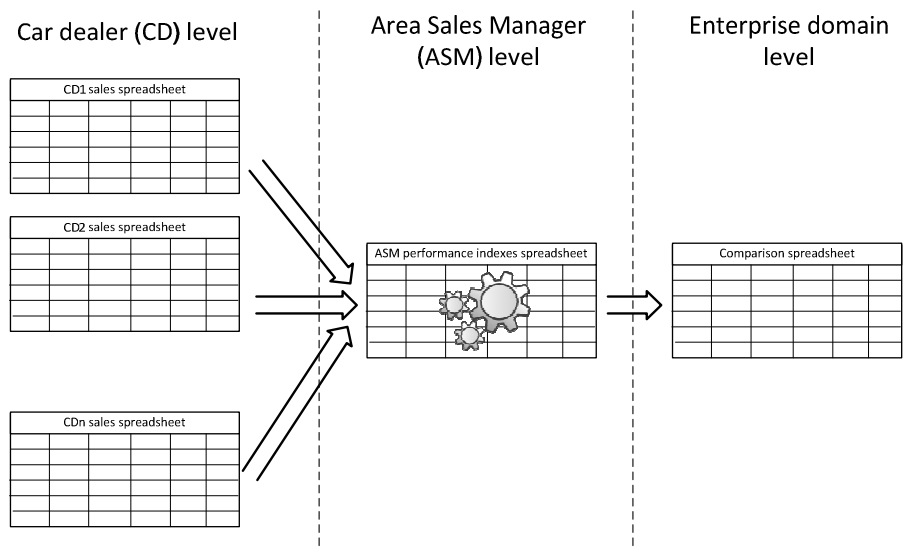

**Figure 1: Car selling enterprise scenario.**

This scenario identifies some problems the cars selling enterprise must face:

- the CDs and the ASM are widely distributed and potentially large in number, therefore there is a strong need for an automatic way of syncing information among them: the CDs should export only the information needed by the ASM in an easy way, avoiding to transfer their entire spreadsheet or communicate changes manually (via email and/or shared storage resources);

- the information exchanged between each CD and the ASM must not be viewed by the other CDs, only the final comparison is visible to everyone, therefore a permission manager system is needed;

- the three levels in Figure 1 form a "distributed spreadsheet chain", since spreadsheets belonging to different levels and actors are linked through shared data, which must be kept always updated;

- every element in the distributed spreadsheet chain should not be aware of belonging to that chain, but it should work as a stand-alone unit relying on a mechanism automatically keeping the chain alive.

## 2.2    Requirements for a Spreadsheet Composition Approach

From a generalization of such a scenario we derive the following requirements for a distributed spreadsheet composition approach and for the supporting ICT tools/platforms:

- support of End-User Computing: being widely used and well known by end users, spreadsheet-based tools can be used as enablers for the creation of Distributed Spreadsheet Compositions, so that no programming skills are required for the composition to take place;

- support of reuse of already existing data: enterprise users often use spreadsheets to store information and/or data extracted from other company's software systems (e.g., ERP system). Thus it is important to provide a tool that allows to combine data belonging to different spreadsheets easily and friendly;

- compliance with the distributed and hierarchical structure of enterprises: the platform is supposed to manage the case in which the spreadsheets belonging to a composition are stored in different machines (e.g., different units of the same company may be geographically distributed). Moreover, the Spreadsheet Composition platform must be aware of the different roles of the employees inside the company in order to apply the correct access rights according to the hierarchical structure of the company;

- support of automatic updates and "Always-on" distributed Composite Spreadsheets: the maintenance of complex relationships over distributed spreadsheet composition is necessary to preserve data consistency. In particular the system is supposed to synchronize the linked spreadsheets automatically even when one or more components of the distributed spreadsheet are offline.

## 3 NOVELTY OF THE APPROACH: PLATFORM FEATURES AND BENEFITS

### 3.1 Approach and Platform Features

The proposed approach to distributed spreadsheet data composition is supported by an IT platform named DISCOM (Distributed Spreadsheet COMposition) based on a client-server architecture: one or more clients interact with the server platform across a network, which can be either an Intranet, a VPN or the Internet.

- The **client** part is a Plug-in module integrated in the client's local spreadsheet application. The module interacts with the composition platform by means of Web Services and allows the user to configure itself and make exportations and importations of spreadsheet cells by means of a graphical interface.
- The **DISCOM platform** exposes interfaces that support the execution of spreadsheet data exportation and importation, manage user accounts and synchronize data across dependent.

Figure 2 illustrates the system operation. Spreadsheet A exports a cell range (dotted area) and Spreadsheet B imports the same range. In the same way, Spreadsheet B can export a range (dashed area) that is imported by Spreadsheet C and so forth. In the above example, Spreadsheet A is an **Exporter**, Spreadsheet C is an **Importer**, and Spreadsheet B is both an importer and an exporter.

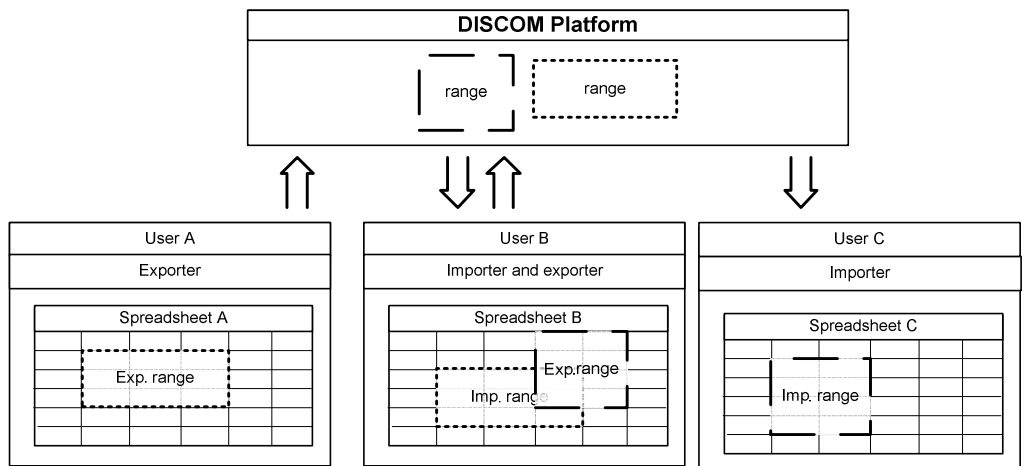

**Figure 2: Spreadsheet composition operation.**

Data propagation is implemented through the following mechanisms:

- the **exporter contribution update** is periodically performed by each exporter Plug-in towards the platform. The Plug-in checks the spreadsheet for exported range modification and refreshes the range image on the platform. In this way, the platform always contains the latest update of the exported ranges. If the user computer is offline, the Plug-in caches the modifications, and exports them to the platform as soon as possible;

- the **importer spreadsheet update** is periodically performed by the importer Plug-in. The Plug-in periodically polls the platform for new contributions or updates. As soon as a new contribution or update is detected the Plug-in imports the contribution and inserts it in the spreadsheet. In this way, the spreadsheet always contains the latest update and updates are propagated in real time.

If a user spreadsheet is both importer and exporter, and at least an exported cell is a function of an imported cell (via direct inclusion or formulas) we call it an **Intermediate** spreadsheet (in our Figure 2 example, Spreadsheet B is an intermediate spreadsheet via direct inclusion of the dashed range in the dotted range). Intermediate spreadsheets allow building spreadsheet chains.

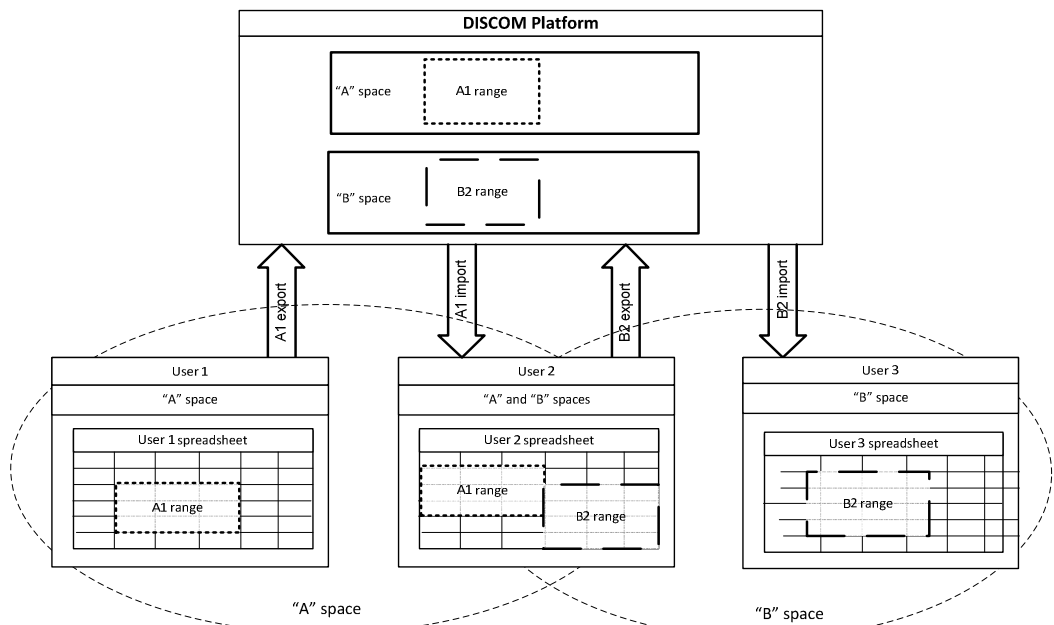

**Figure 3: Example of spaces.**

As we assume that a user personal computer can be switched-off in every moment, a scenario that includes one or more intermediate spreadsheets presents a crucial issue: if an intermediate spreadsheet computer is offline (e.g., switched off or not connected to the network), the data updates cannot propagate through the chain. In order to overcome this issue, the system should support a feature, namely the **data propagation function,** which works as follows:

- when a client Plug-in realizes that the spreadsheet is intermediate it uploads the entire spreadsheet on the platform. Every time an intermediate spreadsheet is updated the Plug-in performs a new upload;
- when an intermediate spreadsheet is offline, the platform runs a local spreadsheet engine in order to recalculate the exported ranges based on fresh import ranges.

In the example of Figure 2, Spreadsheet B is an intermediate spreadsheet and is therefore uploaded on the platform in order to assure data updates from A to C.

The visibility of data exported to the platform can be controlled, allowing data to be exported to all users or part of them. In order to provide more flexible functions to control data lifecycle and visibility, the concept of **Space** has been introduced. A Space is a collection of platform users, one of them being the Space Creator, while the others are either data exporters or data importers.

The Space is a dynamic entity: a Space can exist on the platform to manage a spreadsheet based collaboration carried on by different actors in a limited time period (e.g., "2011 First Quarter Balance Assessment"). A space can also be created to manage spreadsheet based collaborations for a limited set of actors belonging to a well defined unit in an organization.

## 3.2 Platform architecture and implementation

The platform architecture is depicted in Figure 4: the two main platform components are the Plug-in Client and the DISCOM Platform. We adopted an architecture based on a central composition platform in order to provide a central repository that performs automatic synchronization even when one or more contributors are offline.

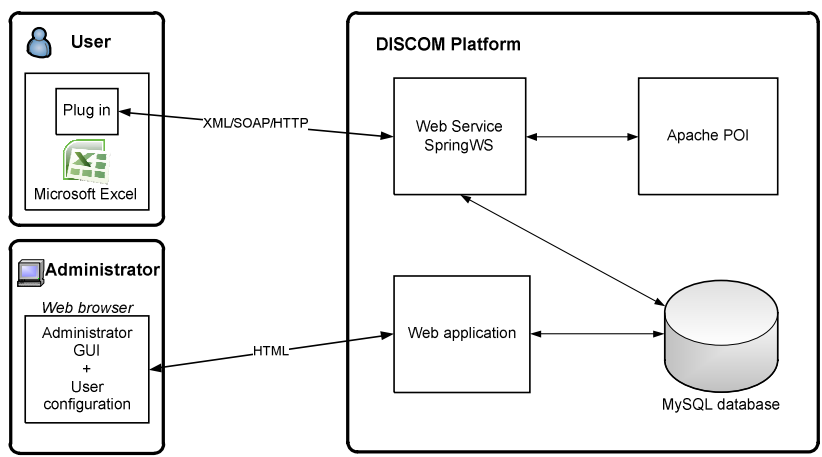

**Figure 4: Platform architecture.**

## Client Plug-In

The Plug-in Client is written in C# using Visual Studio Tool for Office [8], developed for 2003 and 2007 Microsoft Excel versions. It provides a graphical interface that allows users to control the import/export operations; the interface maintains the same look and feel as the other Microsoft Excel functionalities, with respect to different versions of UI layout (command bar or ribbon interface). The main design objective was to integrate it as seamlessly as possible in Excel applications, letting users concentrate on normal Excel commands without worrying about other tools.

The Plug-in interacts with the DISCOM Platform through Web Services technology [9], as Web Services are widely developed and supported in enterprise environments. In particular we adopt standard SOAP messaging. The cell range to be exported is converted to a custom XML document and wrapped in a SOAP request which is sent to the DISCOM Platform. The DISCOM Platform module stores the XML translation of the exported cells in the database. Additionally, the Plug-in stores metadata locally on the client machine:

- exportations and importations information are kept in the spreadsheet, inside file custom properties;
- user identification, server address and other plug-in configuration information are kept in user's local space on the filesystem.

In order to perform the data propagation function, if the Plug-in realizes that the spreadsheet contains an automatic exportation dependent on an automatic importation, it uploads the entire spreadsheet on the platform. Otherwise, only cell ranges are exchanged.

**Server Side Platform**

The server side of the platform is written in Java, using Spring framework [10] as the engine for managing Web Services and reacting to HTTP and SOAP requests. The main interface exposed to the Client Plug-In is a Web Service which provides a set of functions that allow clients to authenticate themselves and perform CRUD (Create, Read, Update , Delete) operations on exported and imported data. It serves client SOAP requests by invoking Spring beans interacting with a MySQL database [11] in a transactional environment.

The data propagation function is realized by a Spring bean which perform this sequence of operations:

1) detection of the intermediate uploaded file to be re-evaluated;
2) insertion of updated values in its automatic importations;
3) recalculation of cell formulas;
4) update of automatic exportations (those modified by the recalculation).

Every step is performed using Apache POI library [12], a set of objects able to open and manipulate Microsoft Excel file directly within Java code.

For administration purposes a web application is provided by the platform. Through this component an administrator can manage user accounts and control the platform correctly working.

The platform can be deployed in two different ways:

- as an enterprise service, creating a "ad-hoc" server inside an enterprise domain to give the enterprise full control over the system and limit spreadsheet data circulation within the enterprise domain;
- as a cloud service hosted on a IaaS (Infrastructure-as-a-Service) provider like Amazon Machine Image – AMI on the Amazon EC2 Infrastructure [13], to enable an easier and quicker setup for spreadsheet compositions which don't need strong control and data flowing-over-the-internet limitations.

## 3.3    The platform at work

In this section we explain how the DISCOM platform can solve the issues presented above using the car selling enterprise scenario presented in section 2.1 as a reference scenario.

Each Car Dealer (CD) and the Area Sales Manager (ASM) are requested to install the plug-in in their Microsoft Excel local applications and should have access to the platform installed in the enterprise domain (or to the platform located in the Cloud).

All the CDs agree with ASM to structure the spreadsheet information using a common format. As an example we assume that the format is 4 columns containing car model, number of sales, average price and total income. Figure 5 depicts the export operation: every CD, through the plug-in interface, assigns its data a name and a description and makes such data available only to the ASM.

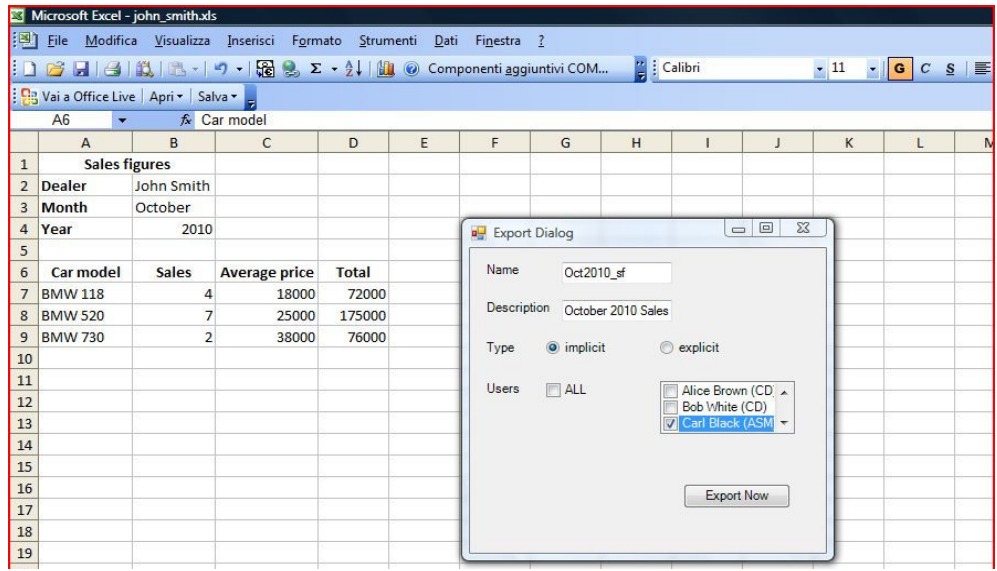

**Figure 5: Cell range exportation. With reference to the car selling enterprise scenario presented in section 2.1, car dealer *John Smith* is exporting Oct. 2010 sales figures to ASM *Carl Black*.**

The ASM can import the CD contributions using the plug-in interface and choosing among the accessible data (see Figure 6). Then it sets the rules (spreadsheet functions) to calculate the performance indexes starting from the imported data, without worrying about the actual data availability. All the CD exportations and the ASM importations are performed automatically, so the plug-in and the platform work in background to keep the data constantly updated to the last changes made by the CDs. To publish the final CDs comparison based upon the calculated performance indexes, the ASM exports it the same way as the CDs do, with the difference that in this case it makes the data available to the entire Space, which represents its area and includes itself and the CDs.

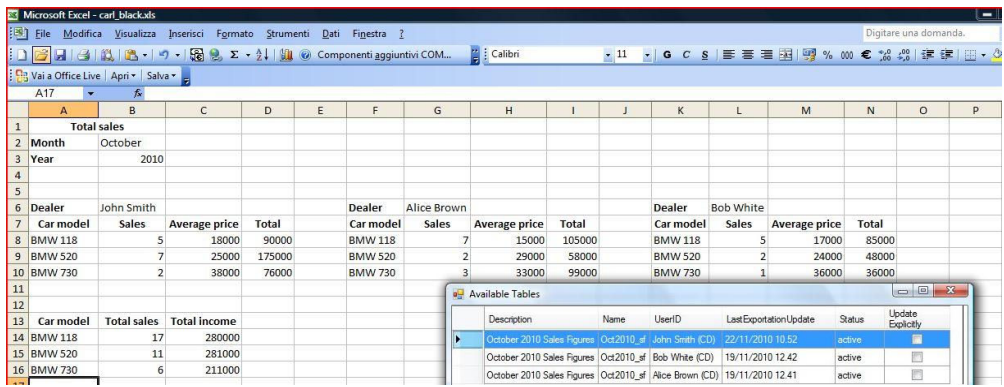

**Figure 6: Cell range importation. With reference to the car selling enterprise scenario presented in section 2.1, ASM *Carl Black* is importing sales figures from three car dealers.**

Since the ASM local spreadsheet is an intermediate spreadsheet (see section 3.1), the platform requires ASM to upload an entire copy of it in order to keep the distributed spreadsheet chain (from CDs exported data to final comparison) always updated. Therefore any CD who wants to view the comparison always views the latest performance indexes even if the ASM is not working on its spreadsheet: the platform has its own copy, knows its dependencies (importations from CDs) and can recalculate it in any moment, publishing the updated final comparison in place of the ASM.

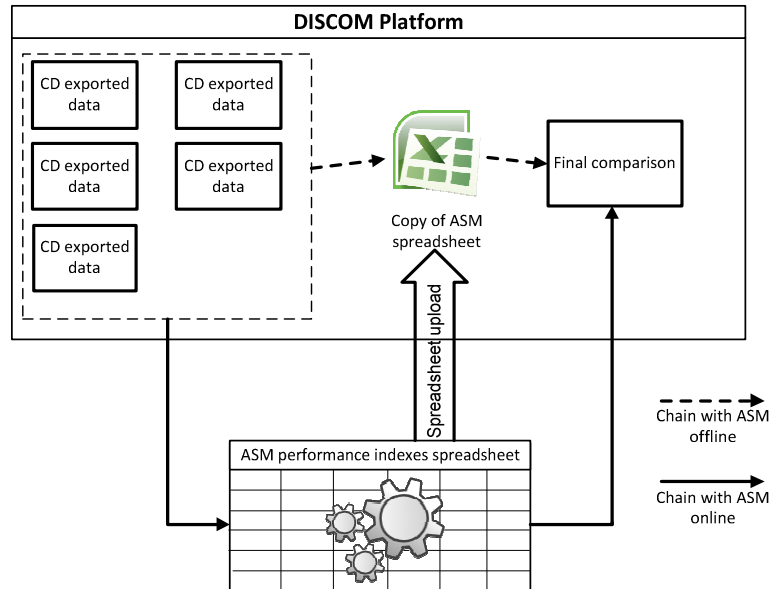

**Figure 7: How the platform manages the chain in car selling enterprise scenario presented in section 2.1**

## 4 COMPARISON WITH OTHER APPROACHES AND TECHNOLOGIES

### 4.1 Other approaches and technologies for Distributed Spreadsheet Composition

Several approaches and solutions are already available for Distributed Spreadsheet Composition: in particular, we will focus on Google Docs [14], on Microsoft SharePoint [15], and on the Data Mashup approaches.

Google Docs is a collaborative tool allowing the sharing of document files among users [14] . It is implemented following the Software-as-a-Service, SaaS, Cloud Computing model according to which users can manipulate spreadsheet files by accessing a Web Application instead of installing software like Microsoft Excel, Open Office, etc. on their own pc. With Google Spreadsheets, users can create, edit, and share spreadsheets without taking care about how and where these files are located because they are stored "in-the-cloud". Moreover, this tool allows users to work simultaneously on the same spreadsheet (i.e., colored cursors will be assigned to the users where each color represents a different user working on the file). Finally, Google Spreadsheets presents the "Cross-workbook references" feature which enables the interconnection among spreadsheets to create the chain of spreadsheets described in Section 3.

One of the drawbacks related to the usage of this tool is that users are supposed to learn how to use a new tool that, even if it is very similar to the one they are already able to use (e.g., MS Excel), it is slightly different (on the contrary the system described in section 3 represents just a new MS Excel feature to the user who can continue using the same tool). Since Google Spreadsheets provides fewer features compared to MS Excel, there might be some compatibility issues between existing MS Excel-created files and the Google Spreadsheets counterpart. A similar tool is proposed in Microsoft Office Web Apps [16] that extends the basic functionalities of Microsoft Office suite products (like the spreadsheet software Excel) to the cloud. In particular, the interaction model is similar to the one provided by Google Docs, allowing multiple users to edit an online spreadsheet using the browser.

Microsoft SharePoint is a suite of enterprise collaboration products, developed by means of the integration of SaaS technologies and client software, providing a large set of functionalities such as website visual composition, enterprise social networking, document sharing, content management, advanced search support and business intelligence applications. In particular, Microsoft SharePoint offers a *Excel Service*, that allows users to publish an Excel spreadsheet on a SharePoint server, thus allowing the sharing of the spreadsheet among user in a workgroup. The spreadsheet can be read and modified using Microsoft Excel software, using a browser and a web application as in Microsoft Office Apps service, or modified by means of Web Service calls.

Spreadsheet hosted in Google or Microsoft services can be accessed also through a set of APIs [17], thus allowing the creation and the maintenance of spreadsheets in a programmatic way (i.e., a programmer can develop an application that retrieves and/or updates cells' values without any intervention by human beings). This feature can be used to exploit the system as the technology standing behind the server-side platform introduced in section 3 because it might cover the role of the component storing the exportation/importation information as well as keeping the chains updated thanks to the "Cross-workbook references" feature. Although this approach is very similar to the one proposed in this paper, it shows the following drawbacks:

- it lacks flexibility because the functionalities provided by the server-side component are limited to those provided by the platform, whereas the server-side component described in section 3 has been developed from the scratch thus it can be easily extended;
- it provides a rigid scheme for security management (see[17]**Error! Reference source not found.**) which represents one of the most important features of the system described in section 3.

Another approach proposed in the literature for the composition of spreadsheets is that of Data Mashups [7] (i.e., applications combining information retrieved from mixed sources and possibly provided in mixed formats). The distinctive feature of products like Yahoo!Pipes [18], Apatar [19], etc. is their focus on the management of different data formats (e.g., spreadsheets, databases, XML files, CSV files, plain text, etc.) and not on the creation/maintenance of distributed spreadsheets continuously kept up to date. Some of the tools supporting Data Mashups might provide a spreadsheet-based user interface in order to exploit the well known paradigm based on grids and cells which is widely used by people all around the world (see [6] for a survey of the available solutions). Although Data Mashup platforms might be used to solve (some of) the issues described in section 1, they have been designed for other purposes (e.g., merging data provided by means of different formats) thus they are not optimized to manage the composition of spreadsheets. In addition Data Mashup tools often require the installation of a new software on the user's pc thus there is - again - the problem related to the fact that a user is supposed to learn how to use a new tool. Finally, the most of the available Mashup tools don't provide capabilities supporting the security thus data stored in spreadsheets might be stolen or changed by malicious behaviors.

## 4.2    Benefits and comparison with other approaches

When compared with the traditional practices for distributed spreadsheet composition, where spreadsheet data composition takes place with almost manual techniques like cross-spreadsheet cell references, manual copy-paste operations, email spreadsheet circulation or single shared spreadsheet location, the proposed approach introduces a different way to compose spreadsheets, removing some or all of the limits and risks which affect the traditional practices.

In particular:

- there is no need for storing spreadsheets in a shared location: after data have been exported to the composition platform, these are available for importation for every user of the platform which is allowed to do it;
- data importations are performed with the assistance of the specific platform Plug-in component, reducing the risks for incorrect copy/paste operations;
- the platform eliminates the need for spreadsheet circulation by email or other means, since data are stored in a centralized server and kept up to date by the platform: whenever a user updates the data in a spreadsheet area which has been "exported" to the platform, these are pushed to the platform server in an automatic way;
- only a single (centralized) copy of the data is maintained, thus eliminating the risk of multiple data sources;
- moreover, the data are updated by the platform even when one of the actor participating to the data workflow is not available (e.g. he/she does not have Excel running on her desktop).

When compared with other approaches to spreadsheet composition, the proposed platform exhibits the interesting following features:

- the platform supports spreadsheet based data composition and aggregation in a hierarchical and distributed way, which is compliant with the usual collaboration practices for distributed spreadsheet management in a complex organization or enterprise;
- the target of the composition is itself a spreadsheet, within which data are kept up to date in real time thanks to the tools provided by the platform, in a transparent way for the end user;
- data updates take place even whenever one of the actor participating in the distributed spreadsheet elaboration process is not available or is not running the spreadsheet application from his desktop;
- no specific software programming skills are required for adopting the approach, since data composition takes place through the usage of the well known tools and methods provided by the traditional spreadsheet editing tools: the proposed approach encourages "end user computing", while providing at the same time a controlled environment when it can take place;
- the proposed platform easily integrates with the usual tools for spreadsheet management and at the same time extends them for supporting collaborative and distributed processes in spreadsheet elaboration in a non invasive way within an organization or industry.

## 5    CONCLUSIONS AND FUTURE DEVELOPMENTS

The DISCOM Platform is an ongoing project: platform and methodology on-the-field trials and experimentations are currently going on in order to evaluate the platform features and usability of the approach. At the moment, an experimental prototype has been installed and is currently used at the Regional Branch of the ICT Department of the Italian Ministry of Justice (CISIA), in order to assess and enhance the platform features and to evaluate the usability of the approach. Several improvements are already under investigation, among which it's worth mentioning:

**Data Modeling and Validation**: thanks to the fact the spreadsheet data are internally represented as XML documents on the platform server, it will be possible to exploit XML Schema Validation [20] in order to provide modeling tools for server side compliance enforcement every time data are exported from the spreadsheet to the platform. This feature will help to greatly reduce format and value errors in spreadsheet cells.

**Data versioning**: as a future work the platform will be equipped with a versioning mechanism, in order to maintain data exportation changes history. Users will be allowed to browse different versions during the importation process.

**ECM integration**: since we are aware that many complex organization runs Enterprise Content Management systems, future research direction will investigate how the proposed approach can fit into the current way ECM are used within the enterprises, and also how the proposed platform will contribute to the ECM features or will benefit from a strong server side integration with them.

More information on the DISCOM platform is available at the address http://www.m3s.it/discom.